\documentclass[aps,pra,showpacs,twocolumn]{revtex4}
\AtBeginDvi{}
\usepackage{graphicx}
\usepackage{epsfig}
\usepackage{dcolumn}% Align table columns on decimal point
\usepackage{bm}% bold math
\def\bra{\langle}
\def\ket{\rangle}

\def\sp{\hat{s}}
\begin{document}

\title{Dissociation dynamics of resonantly coupled bose-fermi mixtures in an optical lattice}

\date{\today}

\author{Takahiko Miyakawa and Pierre Meystre}
\affiliation{Department of Physics, The University of Arizona,
Tucson, Arizona 85721, USA}

\begin{abstract}
We consider the photodissociation of ground-state bosonic
molecules trapped in an optical lattice potential into a
two-component gas of fermionic atoms. The system is assumed to be
described by a single-band resonantly-coupled Bose-Fermi Hubbard
model. We show that in the strong fermion-fermion interaction
limit the dissociation dynamics is governed by a spin-boson
lattice Hamiltonian. 
In the framework of a mean-field analysis
based on a generalized Gutzwiller ansatz, we then examine the
crossover of the dissociation from a regime of independent
single-site dynamics to a regime of cooperative dynamics as the
molecular tunneling increases. 
We also show that in the limits of weak and strong intersite
tunneling the mean-field solutions agree well with the results
from the quantum optical Jaynes-Cummings and Tavis-Cummings models, respectively.
Finally, we identify two types of self-trapping transitions, a {\it coherent}
and an {\it incoherent} one, depending on the ratio of the repulsive
molecule-molecule interaction strength to molecular tunneling.
\end{abstract}


\maketitle

\section{Introduction}
The formation of molecules by Feshbach resonance~\cite{FR} and by
two-photon Raman photoassociation~\cite{PA} from ultracold atomic
gases trapped in an optical lattice potential offers remarkable
opportunities to study some of the long-standing questions of
condensed matter physics in systems under exquisite control, and
is also of much interest for quantum information
science~\cite{Jaksch05,Bloch05}. External control parameters such
as laser and magnetic field strengths can be controlled precisely
in real time, allowing the experimental investigation of the
dynamics of strongly correlated systems from the adiabatic to the
sudden regime, the study of the formation of long-range order and
defects as the system transits across quantum critical points, and
the engineering of exotic many-body states starting from various
quantum phases~\cite{DQPT,DBHM}.

The creation of molecules from cold atoms in optical lattices has
been demonstrated experimentally by several groups. Applying
two-photon Raman lasers on $^{87}$Rb atoms in a 3-dimensional (3D)
lattice, T.~Rom {\it et al.} succeeded in controlling both the
internal and center-of-mass states of the molecules by changing
the detuning of the photoassociation laser fields~\cite{Rom04}.
Unfortunately, the molecular decay arising from the spontaneous
emission induced by Raman laser photons exceeded the
photoassociation coupling, so that this experiment could not
observe coherent Rabi oscillations between atoms and molecules.
The first lattice experiments exploiting the passage of fermions
($^{40}$K atoms) across a Feshbach resonance were performed by
M.~K${\rm \ddot{o}}$hl {\it et al.}~\cite{Kohl}. They observed the
occupation of atoms in higher Bloch bands near the resonance,
indicating that the inclusion of inter-band coupling is crucial
for the description of a broad Feshbach resonance in optical
lattices~\cite{Ho06}. They also created Feshbach molecules from
fermionic atoms in a three-dimensional optical
lattice~\cite{Stoferle}, and found that the measured binding
energy of the molecules is in good agreement with the theoretical
predictions of single-channel models of atoms in a tight
trap~\cite{Busch98,Blume02}. We also mention a recent experiment
by G.~Tialhammer {\it et al.}~\cite{Grimm06}, who created
long-lived $^{87}$Rb$_2$ molecules, with lifetimes of up to
$0.7$s, in a three-dimensional optical lattice via a narrow
Feshbach resonance.

In a recent experiment, C.~Ryu {\it et al.} generated a quantum
degenerate gas of molecules via photoassociation of $^{87}$Rb
atoms, and observed a clear coherent atom-molecule oscillation in
an optical lattice, with Rabi frequency $\approx
4.13$kHz~\cite{Ryu05}. In this experiment the atoms were initially
in the Mott-insulator (MI) phase. For the duration of the
photoassociation pulse each atom was trapped in the ground state
of a single lattice site with trapping frequency $\omega_{\rm
lattice}\approx 2\pi\times 27$kHz, and the atom-atom interaction
strength and atomic intersite tunneling were of the order of $\sim
10$kHz and $\lesssim 10$Hz, respectively. In general, the initial
atomic state in that experiment can be in either the MI or the
superfluid (SF) phase, depending on the depth of the optical
lattice. Combined with the interplay between two-body
interactions, intersite tunneling, and atom-molecule coupling,
this results in the photoassociation process exhibiting a very
rich dynamics.

Motivated by these experimental advances, this paper considers a
gas of two-component fermionic atoms coupled to bosonic molecules
via photoassociation in an optical lattice. The system consists
initially of bosonic molecules only, assumed to be in a ground
state corresponding to either a MI phase or a SF phase, and a cw
photoassociation beam is switched on instantaneously at $t=0$. Our
goal is to study the subsequent dissociation dynamics of the
bosonic molecules into fermionic atoms. We examine the crossover
of the dynamics from an independent single-site regime to a
cooperative regime. We also show the appearance of two types of
self-trapping transitions ~\cite{Milburn97,Smerzi97}, which we
characterize as {\it coherent} and {\it incoherent}, as the ratio
of repulsive molecule-molecule interaction strength to molecular
tunneling is varied.

Section II discusses our model and formulates it as a single-band
resonantly coupled Bose-Fermi Hubbard model. We show that in this
model the dissociation dynamics can be described by
a spin-boson lattice Hamiltonian in terms of pseudo-spin formalism.
In this section we also give a formulation of mean field
approximation based on a generalized Gutzwiller (GW) ansatz. Section III
discusses the dissociation dynamics in several limiting cases.
Specifically, we show that in the weak and strong molecular
tunneling limits, respectively, it can be described in terms of
generalized versions of the Jaynes-Cummings and Tavis-Cummings
models of quantum optics. Section IV then presents numerical
results of the GW mean-field dynamics, and section VI is a summary
and outlook.

\section{Model}
\subsection{ Spin-Boson lattice Hamiltonian}

We consider at zero temperature a gas of fermionic atoms of mass
$m_f$ and spin $\sigma=\uparrow,\downarrow$ trapped in an optical
lattice potential. The fermionic atoms can be coherently combined
into bosonic molecules of mass $m_b=2m_f$ via two-photon Raman
photoassociation. The lattice lasers are adjusted in such a way
that the tight binding approximation is well justified. Since in
lattice photoassociation experiments the atom-molecule interaction energy is
typically small compared to the trapping energies of the atoms
and molecules~\cite{Rom04,Ryu05} (note that this is not so for the
broad Feshbach resonance of $^{40}$K atoms~\cite{Kohl,Stoferle}) the
system can be described by the single-band resonantly coupled
Bose-Fermi Hubbard model~\cite{Dickerscheid,Carr05,Zhou05,Miyakawa06}
\begin{eqnarray}
\label{BFH}
\hat{H}_{BF}=\hat{H}_f+\hat{H}_b+\hat{V}_{bf},
\end{eqnarray}
where
\begin{eqnarray}
\label{FHubbard}
\hat{H}_f&=&-\hbar J_f\sum_{\langle ij \rangle\sigma}(\hat{f}_{i\sigma}^\dagger
\hat{f}_{j\sigma}+{\rm H.c.})
+\hbar\omega_f\sum_{i\sigma} \hat{n}^f_{i\sigma}\nonumber\\
&+&\hbar U_f\sum_{i} \hat{n}^f_{i\uparrow}\hat{n}^f_{i\downarrow},\\
\label{BHubbard}
\hat{H}_b&=&-\hbar J_b\sum_{\langle ij \rangle}(\hat{b}_{i}^\dagger
\hat{b}_{j}+{\rm H.c.})
+\hbar(\omega_d+\omega_b)\sum_{i} \hat{n}^b_{i}\nonumber\\
&+&\frac{\hbar U_b}{2}\sum_{i}\hat{n}^b_i(\hat{n}^b_i-1),\\
\label{BFcoupling}
\hat{V}_{bf}&=&\hbar U_{bf}\sum_{i\sigma}\hat{n}^f_{i\sigma}\hat{n}^b_i
+\hbar g\sum_i(\hat{f}^\dagger_{i\uparrow}\hat{f}_{i\downarrow}^\dagger\hat{b}_i+{\rm H.c.}).
\end{eqnarray}
Here $\hat{f}_{i\sigma}$ and $\hat{b}_i$ are annihilation
operators for the fermionic atoms of spin $\sigma$ and bosonic
molecules at the $i$-th site, respectively. The corresponding
number operators $\hat{n}^f_{i\sigma}$ and $\hat{n}^b_i$ have
eigenvalues $n^f_{i\sigma}$ and $n^b_i$. The conserved quantity
\begin{equation}
N_{F}^{\rm tot}=2\sum_i n^b_i +\sum_{i\sigma} n^f_{i\sigma}
\end{equation}
is the total number of fermions in the system.

The Hamiltonians $\hat{H}_f$ and $\hat{H}_b$ describe standard
Hubbard models for fermions and bosons, respectively, and
$\hat{V}_{bf}$ describes  the interactions between fermions and
bosons. The terms proportional to $J_{\alpha}$, where
$\alpha=f,b$, account for the tunneling of particles between
nearest neighbor sites denoted by $\langle ij \rangle$. The
single-particle center-of-mass energies are $\hbar\omega_\alpha$,
and the effective detuning between the Raman lasers and the
difference in internal energies of the atoms and molecules is
$\hbar\omega_d$. The on-site fermion-fermion, boson-boson, and
boson-fermion collisions are described by the interaction
strengths $U_f$, $U_b (>0)$, and $U_{bf}$, respectively. Finally,
the term proportional to $g$ in $\hat{V}_{bf}$ describes the
conversion of fermionic atoms into bosonic molecules, and vice
versa, via photoassociation.

We concentrate on the case where the system consists initially of
bosonic molecules only, and the corresponding initial state is the
ground state of the Bose-Hubbard Hamiltonian $\hat{H}_b$. It is
well-known that the ground state of this model describes either a
MI phase or a SF phase, depending on
the ratio $U_b/zJ_b$, where $z$ is a number of nearest neighbor
sites, for a fixed number of
bosons~\cite{Fisher,Rokhsar91,Krauth92,Sachdev,Jaksch98,Greiner02}.

\begin{figure}
    \begin{center}
      \includegraphics[width=7.5cm,height=5cm]{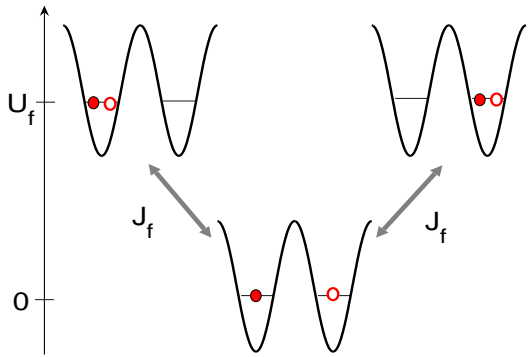}
      \end{center}
    \caption{(Color online) Schematics of the tunneling of a Fermi pair in the limit
      $U_f\gg zJ_f$. Filled and open circles present fermions with different
      spins.}
    \label{pairtunnel}
\end{figure}

In the following we consider a strongly confining regime
$|U_f|\gg zJ_f$. In this regime, Fermi pair states, that is, pairs
of fermions with spins up and down occupying the same site, see
upper diagrams in Fig.~\ref{pairtunnel}, have an energy separation
$U_f$ compared to unpaired states such as in the lower diagram.
This allows us to treat the fermionic tunneling that couples the
paired and unpaired states perturbatively. Since in addition the
photodissociation of a molecule always creates a Fermi pair at a
given lattice site, the unpaired states can then be adiabatically
eliminated and integrated out. This results in the tunneling of
Fermi pairs via those virtual states (Fig.~\ref{pairtunnel}), and
the Hamiltonian $\hat{H}_f$ is mapped on a pseudo-spin-1/2 system
(XXZ model) by second-order degenerate perturbation
theory~\cite{Emery76,Carr05},
\begin{equation}
\label{XXZ} \hat{H}_s=\hbar\omega_{s}\sum_{i}(2\sp_{fi}^z+1)
+\frac{\hbar J_s}{2}\sum_{\langle ij \rangle}(\sp_{fi}^+\sp_{fj}^-+
\sp_{fi}^-\sp_{fj}^+-2\sp_{fi}^z\sp_{fj}^z).
\end{equation}
Here $\omega_s=\omega_f+U_f/2$ and $J_s=4J_f^2/U_f$, and we have
introduced the pseudo-spin operators describing the creation and
annihilation of a Fermi pair
\begin{equation}
  \sp_{fi}^+=\hat{f}^\dagger_{i\uparrow}\hat{f}^\dagger_{i\downarrow},
  \quad
  \sp_{fi}^-=\hat{f}_{i\downarrow}\hat{f}_{i\uparrow}\,
  \quad
\end{equation}
and
\begin{equation}
  \sp_{fi}^z=\frac{1}{2}(\hat{n}^f_{i\uparrow}+\hat{n}^f_{i\downarrow}-1)
\end{equation}
at the $i$-th site~\cite{Search03}. This mapping amounts to
describing fermionic pairs as ``effective two-level atoms'' whose
upper level corresponds to a pair of fermions, and lower level the
absence of such a pair. Thus the presence of paired states at the
$i$-th site is described by the spin-up state $|s^z_{fi}=1/2
\rangle$ and its absence by the spin-down state $|s^z_{fi}=-1/2
\rangle$. The pseudo-spin operators obey the commutation relations
$$
[\sp_{fi}^z,\sp_{fj}^\pm]=\pm\sp_{fi}^\pm\delta_{i,j},\quad
[\sp_{fi}^+,\sp_{fj}^-]=2\sp_{fi}^z\delta_{i,j}.
$$

In terms of these operators the interaction Hamiltonian $\hat{V}_{bf}$
can be re-expressed as
\begin{equation}
\label{SBC}
\hat{V}_{sb}=\hbar U_{bf}\sum_{i}(2\sp_{fi}^z+1)\hat{n}^b_i
+\hbar g\sum_{i}(\hat{b}^\dagger_i\sp_{fi}^-+\sp_{fi}^+\hat{b}_i).
\end{equation}
We thus conclude that in the strongly confining regime the
dissociation of initially bosonic molecular states into fermionic
atoms is governed by the spin-boson lattice Hamiltonian~\cite{Carr05}
\begin{equation}
\label{SBH}
\hat{H}_{SB}=\hat{H}_s+\hat{H}_b+\hat{V}_{sb}.
\end{equation}
This effective Hamiltonian is valid for both of attractive and
repulsive fermion-fermion interaction cases for the problem at
hand.

\subsection{Gutzwiller mean-field ansatz}
In order to proceed we now introduce a Gutzwiller-type (GW)
variational ansatz~\cite{Rokhsar91,Krauth92} for the many-body wave
function as
\begin{equation}
|\Psi(t)\rangle=\prod_{i=1}^{N_s}
\left(
\sum_{n_i=0}^{\infty}\sum_{s_i=-1/2}^{1/2}f^{(i)}_{n_i,s_i}(t)
|n_i\rangle_b \otimes |s_i\rangle_{fp}
\right),
\end{equation}
where $N_s$ is the number of lattice sites, and $|n_i\rangle_b$ and $|s_i\rangle_{fp}$
refer to number states of the bosons and Fermi pair at the $i$-th site, respectively.
The normalization condition
$\sum_{n_i,s_i}|f^{(i)}_{n_i,s_i}(t)|^2=1$ is satisfied
at every single site.
From the time-dependent variational
principle, the time evolution of the variational parameters
$f^{(i)}_{n_i,s_i}$ is obtained via the minimization
\begin{equation}
\label{TDGA}
\frac{\partial}{\partial f^{(i)*}_{n_i,s_i}}\left\langle \Psi(t)\left|
i\hbar\frac{\partial}{\partial t}-\hat{H}_{SB} \right|\Psi(t)\right\rangle=0.
\end{equation}
The time-dependent Gutzwiller approach has previously been used in
the studies of several ultracold systems on lattices, such as
bosonic atomic-molecular gases~\cite{Jaksch02,Damski03a},
disorder~\cite{Damski03b}, inhomogeneous~\cite{Zakrzewski}
Bose-Hubbard systems and non-resonant Bose-Fermi
mixtures~\cite{Fehrmann}. In all cases the time dependence of the
system parameters was assumed to be slow enough that the system is
allowed to evolve adiabatically.

\begin{figure}
    \begin{center}
      \includegraphics[width=7.5cm,height=5cm]{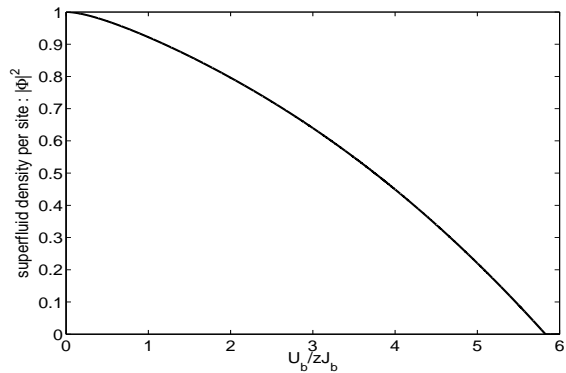}
      \end{center}
    \caption{Superfluid density $|\Phi|^2$ of the molecular ground state
      as a function of $U_b/zJ_b$. The MI-SF critical point is at about $5.83$.}
    \label{gssforder}
\end{figure}

In the present case, the ground state of the bosonic molecules in
the absence of photoassociation is given by the stationary
solution of Eq.(\ref{TDGA}) with $f^{(i)}_{n_i,1/2}=0$ and $g=0$.
One of the merits of the Gutzwiller ansatz is its prediction of
the MI-SF phase transition for a commensurate filling factor. For
a fixed number of bosonic molecules, the ground state is
determined by the single parameter $U_b/zJ_b$ and the
corresponding configuration of $\{f^{(i)}_{n_i,-1/2}\}$ is uniform
over all lattice sites. At $U_b=0$, corresponding to the deep SF
regime, the ground state is a coherent state and the variation
parameters for the system for the filling factor $\nu$ are given
by
\begin{equation}
\label{Poissonian}
f^{(i)}_{n_i,-1/2}=e^{-\nu/2}\frac{(\sqrt{\nu})^{n_i}}{\sqrt{n_i!}},
\end{equation}
leading to Poissonian number statistics at each site. For a finite
$U_b$ the ground-state number statistics of molecules becomes
subpoissonian, and in the MI phase, $U_b/zJ_b \gtrsim 5.83$ for
$\nu=1$, it becomes a Fock state
$f^{(i)}_{n_i,-1/2}=\delta_{n_i,\nu}$. Fig.~\ref{gssforder} shows
the superfluid density per site $|\Phi|^2$ of the molecular ground
state, as a function of $U_b/zJ_b$ for the commensurate case
$\nu=1$, where $\Phi=\langle \hat{b}_i\rangle$.

It can be shown from the minimization of Eq.~(\ref{TDGA}) that the
resulting variational parameters $\{f^{(i)}_{n_i,s_i}\}$
remain uniform over the entire lattice when starting from a
uniform configuration. This is the case for the dissociation of
ground-state bosonic molecules.
In that case the equations of motion for the site-independent parameters
$\{f_{n,s}\}$ are found to be
\begin{eqnarray}
\label{GWMFD}
%i\dot{f}_{n,\sigma}
i\frac{\partial f_{n,s}}{\partial t}
&=& h_{n,s}f_{n,s}+ g\left[
\sqrt{n}f_{n-1,s+1}+\sqrt{n+1}f_{n+1,s-1}\right]\nonumber\\
&-&zJ_b\left[
\sqrt{n+1} \Phi^* f_{n+1,s}+\sqrt{n}f_{n-1,s} \Phi
\right]\nonumber\\
&+&\frac{zJ_s}{2}\left[
\Delta^* f_{n,s+1}+f_{n,s-1}\Delta
-2s M f_{n,s}\right],
\end{eqnarray}
where $h_{n,s}=\delta n+(U_b/2) n(n-1)+U_{bf} n(2s+1)$
and $\delta=\omega_d+\omega_b-2\omega_s$. Here, we have used the
average number conservation
\begin{eqnarray}
  N_F^{\rm tot}&=&2N_m=2\sum_{i}\left(\langle \hat{n}^b_{i} \rangle +
  \langle \hat{s}_{fi}\rangle+\frac{1}{2}\right)\nonumber\\
  &=&2N_s\left\{\sum_{n,s}\left(n+s
  +\frac{1}{2}\right)|f_{n,s}|^2  \right\}
\end{eqnarray}
and $N_m$ is the initial number of molecules. We have also introduced the
superfluid order parameters of the molecular bosons, $\Phi$, and
the atomic Fermi pairs, $\Delta$, and the {\it magnetization} $M$
of the pseudo-spin as
\begin{eqnarray*}
\Phi(t) &\equiv& \bra \hat{b}_i \ket
=\sum_{n,s}\sqrt{n+1}f^*_{n,s}f_{n+1,s},\\
\Delta(t) &\equiv& \bra \hat{f}_{i\downarrow}\hat{f}_{i\uparrow}\ket
=\bra \sp_{fi}^-\ket
=\sum_{n,s}f^*_{n,s}f_{n,s+1},\\
M(t) &\equiv& \bra \sp_{fi}^z \ket
=\sum_{n,s}s\left|f_{n,s}\right|^2.
\end{eqnarray*}

Note that in Eq.~(\ref{GWMFD}), the tunneling of both bosons and
Fermi pairs occurs only via their own order parameters, and
fluctuations about their mean values are ignored. In this sense,
the GW ansatz can be thought of as a kind of mean-field
approximation.

\section{Limiting cases}

Before turning to the presentation of a numerical study of
Eq.(\ref{GWMFD}), we first discuss analytical results for the
dissociation dynamics in several limiting cases that provide us
with some useful intuitive understanding of its behavior.

\subsection{Generalized Jaynes-Cummings dynamics : $zJ_b, zJ_s \ll g$ }

In the limit $J_b, J_s \to 0$, the spin-boson Hamiltonian
$\hat{H}_{SB}$ reduces to a generalized version of Jaynes-Cummings
(J-C) model~\cite{Scully} at each lattice site. This situation can
be solved exactly~\cite{Search03}. The average number of molecules
per site, $n_b$, is
\begin{equation}
\label{JCsolution}
n_b(t)=\sum_{n}p(n) \frac{4g^2n}{\mathcal{R}_{n-1}^2}
\cos^2{\left(\frac{\mathcal{R}_{n-1} t}{2}\right)},
\end{equation}
where the Rabi frequency $\mathcal{R}_n$,
\begin{equation}
\mathcal{R}_n=\sqrt{\{\delta+(U_b-2U_{bf})n\}^2+4g^2 (n+1)},
\end{equation}
depends on the number $n$ of molecules. In Eq.~(\ref{JCsolution}),
$p(n)=|f_{n,-1/2}(t=0)|^2$ is the initial number statistics
of the molecules at each lattice site. We note that the GW mean-field
equation (\ref{GWMFD}) has the same exact solution and thus
describes the correct dynamics in this limit.
\begin{figure}
    \begin{center}
      \includegraphics[width=7.5cm,height=5cm]{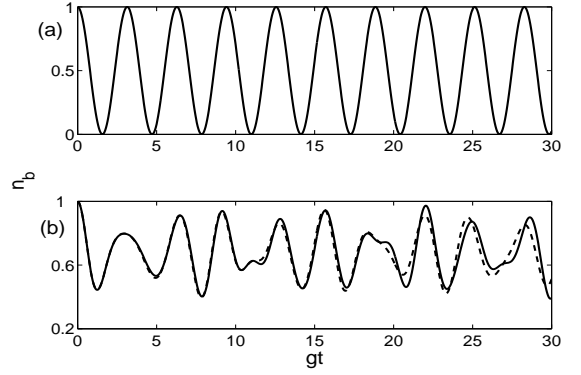}
      \end{center}
    \caption{Time evolution of the average molecule number per site $n_b$
    according to the Jaynes-Cummings dynamics for initially (a) a
    number state corresponding to the Mott-Insulator phase and
    (b) a coherent state.
    Both situations are for $U_{bf}=0$ and
    exact resonance $\delta=0$, and for $\nu=1$.
    The dashed line is the GW mean-field solution for $zJ_b/g=0.1$.}
    \label{JCdy}
\end{figure}

The time evolution of $n_b$ at exact resonance, $\delta=0$ for a
molecular field initially in the MI phase ($U_b\gg zJ_b$) and the
superfluid phase ($U_b=0$) is shown in Fig.~\ref{JCdy}~(a) and
Fig.~\ref{JCdy}~(b), respectively, for the filling factor
$\nu=N_m/N_s=1$. In this example we have taken $U_{bf}=0$. Note
also that the on-site interaction between atoms and molecules
plays no role in the dynamics starting from the MI state for
$\nu=1$. The dashed line in Fig.~\ref{JCdy} (b), obtained from
Eq.~(\ref{GWMFD}) for small but finite tunneling, $zJ_b/g=0.1$,
illustrates the small deviation from the Jaynes-Cummings
solution~(\ref{JCsolution}) in that case.

These results demonstrate that the observation of the number of
atoms, which is the difference between the total number of
fermions and twice the molecule number and is easier to observe
than that of number of molecules, can reveal the precise number
statistics of the molecules and hence their ground-state
properties. A similar argument was previously given by K.
M{\o}lmer in the context of bosonic atom-molecule gases in an
optical lattice~\cite{Molmer03}.

\subsection{Generalized Tavis-Cummings dynamics : $zJ_b \gg g$ and $ U_b/zJ_b\lesssim 1$}

As shown in Fig.~\ref{gssforder}, for $U_b/zJ_b\lesssim 1$ the
initial molecular ground state contains a large condensate
fraction or high occupation of the zero-momentum state. Moreover,
the application of a weak photoassociation coupling, $zJ_b\gg g$,
presumably preserves the coherent character of the molecular
condensate. Owing to Bose enhancement, in that limit the
dissociation dynamics is therefore expected to be dominated by the
condensate of bosonic molecules.

Assuming $J_s=0$ for simplicity, the momentum representation of
the Hamiltonian (\ref{SBH}) is
\begin{eqnarray}
\label{momentumrep}
\hat{H}_{SB}
&=&\sum_{\bm k}
\left[
  \hbar\omega_{\bm k}\hat{b}_{\bm k}^\dagger\hat{b}_{\bm k}
  +\frac{\hbar g}{\sqrt{N_s}}
  \sum_i \left(e^{i{\bm k \bm r_i}} \hat{b}_{\bm k}^\dagger \sp_{fi}^-
  +{\rm H.c.}\right)\right]\nonumber\\
&+&\frac{\hbar U_{bf}}{N_s}\sum_{\bm k_1,\bm k_2}\sum_{i}
  e^{i({\bm k_1-\bm k_2}){\bm r_i}}\hat{b}_{\bm k_1}^\dagger
  \hat{b}_{\bm k_2}(2\sp_{fi}^z+1)\nonumber\\
&+&\frac{\hbar U_b}{2N_s}\sum_{\bm k_1,\bm k_2}\sum_{\bm k_3,\bm k_4}
  \delta_{\bm k_1+\bm k_2,\bm k_3+\bm k_4}
  \hat{b}_{\bm k_1}^\dagger \hat{b}_{\bm k_2}^\dagger\hat{b}_{\bm k_3}
  \hat{b}_{\bm k_4}
\end{eqnarray}
where $\hat{b}_{\bm k}$ is the annihilation operator of the
bosonic molecules with momentum ${\bm k}$ and ${\bm r_i}$ is the
coordinate of site $i$. The single-molecule frequency for a
regular lattice of lattice constant $a$ in $d$ dimensions is given
by
\[
\omega_{\bm k}=\delta-2J_b\sum_{j=1}^d\cos(k_j a).
\]

In the extreme superfluid case, the non-condensed molecule
fraction is negligible and a single-mode approximation that
accounts only for the zero-momentum state is appropriate. The
replacement $\hat{b}_{\bm k} \to \hat{b}_{\bm 0}$, together with
the approximate number conservation
\begin{equation}
\label{appNC}
N_m=\hat{n}_0+\sum_i(\sp_{fi}^z+1/2),
\end{equation}
where $\hat{n}_0=\hat{b}_{\bm 0}^\dagger\hat{b}_{\bm 0}$,
simplifies then the Hamiltonian~(\ref{momentumrep}) to a
generalized version of the Tavis-Cummings (T-C) model of quantum
optics~\cite{Tavis}
\begin{equation}
\label{TCH}
\hat{H}_0\to \hbar\delta_{0}\hat{n}_0
+\frac{\hbar g}{\sqrt{N_s}}\left[\hat{b}_{\bm 0}^\dagger \hat{S}_{f}^-+ {\rm
H.c.}\right]
+\frac{\hbar U_{0}}{2N_s}\hat{n}_0^2.
\end{equation}
Here $\hat{\bm S}_f=\sum_{i}\hat{{\bm s}}_{fi}$ is a collective
spin operator over the entire lattice, and the effective detuning
and effective interaction strength are now
\[
\delta_{0}=\delta-zJ_b+2\nu U_{bf},\quad
U_{0}=U_b-4U_{bf},
\]
where we have neglected a term of ${\mathcal O}(1/N_s)$.
As a result of the absence of the phase factor $e^{i{\bm k \bm
r_i}}$ for the zero-momentum condensate molecules in the
conversion term of $\hat{H}_0$, the system undergoes cooperative
oscillations between atoms and molecules.  Similar models have
been applied to the atom-molecule dynamics across the BCS-BEC
crossover~\cite{BCSBECdy,Uys05} and the dissociation dynamics of a
molecular BEC~\cite{Jack05} in free space.

Although the Hamiltonian (\ref{TCH}) can be solved numerically by
direct diagonalization, it is useful to consider an approximate
analytical solution valid in the semiclassical
limit~\cite{Miyakawa05}. By treating the molecule number operator
$\hat{n}_{0}\to n_0$ classically and taking into account number
conservation~(\ref{appNC}) and energy conservation, we can recast
the Tavis-Cummings dynamics in the form of a Newton equation
\begin{equation}
\label{Newtoneq}
\frac{d^2}{dt^2}n_0(t)=-\frac{d}{dn_0}V(n_0),
\end{equation}
where the potential $V(n)$ is given by~\cite{Miyakawa05}
\begin{eqnarray}
\label{scpotential}
V(n)&=&\frac{U_{0}^2}{8N_s^2}n^4
+\left\{
\frac{2g^2}{N_s}+\frac{\delta_{0}U_{0}}{2N_s}
\right\}n^3\nonumber\\
&-&\left\{
4g^2\left(\nu-\frac{1}{2}\right)-\frac{\delta_{0}^2}{2}
+\frac{\delta_{0}U_{0}\nu}{2}+\frac{U_{0}^2\nu^2}{4}
\right\}n^2\nonumber\\
&-&\left\{
2g^2\left(1-\nu\right)+\delta_{0}^2
+\frac{\delta_{0}U_{0}\nu}{2}
\right\}N_m n.
\end{eqnarray}

In the case $U_{0}=0$, the potential is a cubic function of $n$,
and $n_0(t)$ is given in terms of the Jacobian elliptic function
${\rm sn}(\theta;k)$~\cite{elliptic} as
\begin{equation}
\label{TCsolution}
\frac{n_0(t)}{N_m}=1-(1-n_+){\rm sn}^2\left(\sqrt{(1-n_-)\nu}gt;k\right),
\end{equation}
where
\begin{eqnarray*}
  n_{\pm}&=&\frac{1}{2\nu}
  \left[
    -\left\{
    \left(\frac{\delta_{0}}{2g}\right)^2-\nu+1
    \right\}\right.\\
    &\pm&
    \left. \sqrt{\left\{
      \left(\frac{\delta_{0}}{2g}\right)^2-\nu+1
      \right\}^2
      +4\nu\left(\frac{\delta_{0}}{2g}\right)^2}
    \right],
\end{eqnarray*}
and $k = \sqrt{(1-n_+)/(1-n_-)}$.

For $\delta_0=0$, the classical trajectory approaches an unstable
equilibrium point at $n_0=0$ when $\nu=1$~\cite{Uys05}. It can be
shown from Eq.~(\ref{TCsolution}) that $\nu\neq 1$ excludes the
presence of unstable extremum. For large detunings,
$\delta_{0}/2g\gg \sqrt{\nu}, 1$, the oscillations become
sinusoidal
\begin{equation}
\frac{n_0(t)}{N_m}\to 1-\left(\frac{2g}{\delta_{0}}\right)^2
\sin^2{\left(\frac{\Omega t}{2}\right)},
\end{equation}
where
\begin{equation}
\Omega\approx \sqrt{\delta_{0}^2+4(1+\nu)g^2}.
\end{equation}

\begin{figure}
    \begin{center}
      \includegraphics[height=5cm,width=7.5cm]{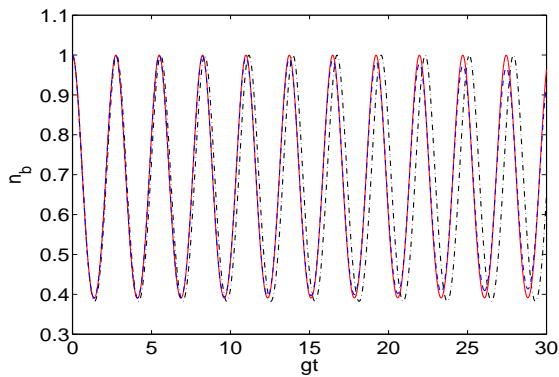}
      \end{center}
    \caption{(Color online) Time evolution of $n_b$
      for $\delta_{0}/g=-1$, $U_b=U_{bf}=0$, and $\nu=1$. The
      solid and dashed lines correspond to the semiclassical and quantum
      ($N_m=2000$) solutions of the Tavis-Cummings dynamics,
      respectively, and the
      dot-dashed line is the GW mean-field solution for $zJ_b/g=20$.}
    \label{TCdy}
\end{figure}

Figure~\ref{TCdy} shows the time evolution of $n_b$ for
$\delta_{0}/g=-1$, $U_b=U_{bf}=0$, and $\nu=1$. The solid and
dashed lines are the semiclassical and quantum-mechanical
Tavis-Cummings solutions, respectively, for $N_m=2000$ and
$n_b=n_0$. The GW mean-field solution (dot-dashed line for
$zJ_b/g=20$) is seen to agree quantitatively with the
Tavis-Cummings dynamics in the limit $zJ_b\gg g$ for
$U_b/zJ_b\lesssim 1$.

\section{Gutzwiller mean-field dynamics}

The preceding section illustrated in limiting cases the crossover
of the dissociation dynamics from a regime of independent
single-site dynamics to a regime of cooperative dynamics as the
molecular tunneling is increased. In this section we examine the
characteristics of the dissociation in these regimes in more
detail within the framework of a Gutzwiller ansatz. We also
demonstrate the appearance of two types of self-trapping
transitions, depending on the ratio $U_b/zJ_b$. We restrict our
discussion to the case $J_s=U_{bf}=0$ and to the commensurate case
$\nu=1$ for simplicity. The effects of the XXZ term proportional
to $J_s$ in the Hamiltonian $\hat{H}_s$ are briefly discussed in
the last subsection.

\subsection{Crossover between single-site and cooperative dynamics}

\begin{figure}
    \begin{center}
      \includegraphics[width=7.5cm,height=5cm]{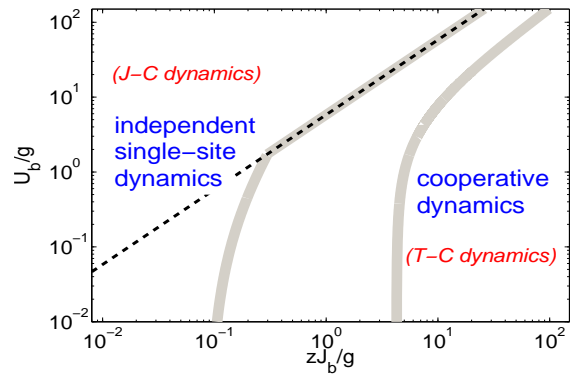}
      \end{center}
    \caption{(Color online) Phase diagram of the GW mean-field dynamics.
      The dashed line indicates the MI-SF phase transition
      for the initial molecular ground state.}
    \label{mfdpd}
\end{figure}

Figure~\ref{mfdpd} shows a schematic phase diagram summarizing the
characteristics of the GW dynamics in the $U_z/g$ - $zJ_b/g$
plane. For $zJ_b\gg g$ and $U_b/zJ_b\lesssim 1$, the dynamics is
dominated by the molecular condensate and hence cooperative. In
the opposite limit, $zJ_b \ll g$, the dynamics on a given lattice
site becomes increasingly independent of the other sites. Note
that the thick gray lines in the figure do not indicate sharp
boundaries between these regimes. The dashed line marks the
separation between initial molecular ground states in the the MI
and SF phases. Note that for a molecular field initially in the MI
phase ($U_b/zJ_b\gtrsim 5.83$), the full Rabi oscillation of
Fig.~\ref{JCdy}a is always recovered at every single site since $\langle
\hat{b}_i \rangle=0$ in that case.

\begin{figure}
    \begin{center}
      \includegraphics[height=5cm,width=7.5cm,clip]{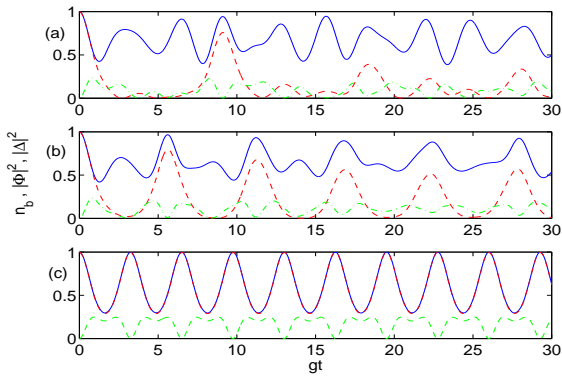}
      \end{center}
    \caption{(Color online) Time evolution of $n_b$ (solid line),
      $|\Phi|^2$ (dashed line),
      and $|\Delta|^2$ (dot-dashed line) for (a) $zJ_b/g=0.1$,
      (b) $zJ_b/g=1$, and (c) $zJ_b/g=10$,
      and for $\delta=zJ_b$, $U_b/zJ_b=0.1$.}
    \label{crossover}
\end{figure}

\begin{figure}
    \begin{center}
      \includegraphics[height=5cm,width=7.5cm,clip]{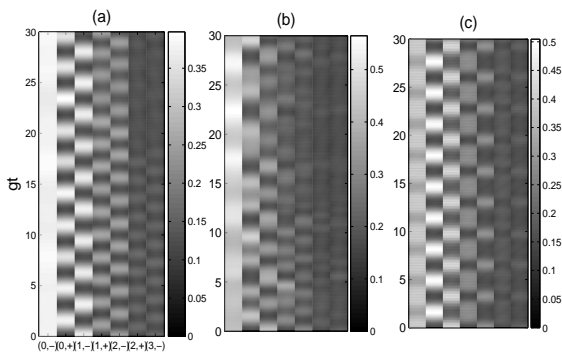}
      \end{center}
    \caption{Gray scale rendition of the time evolution of the probabilities
    $|f_{n,s}|^2$ of the states $(n,sign(s))$ for (a) $zJ_b/g=0.1$,
      (b) $zJ_b=1$, and (c) $zJ_b/g=10$, respectively. The darker
      shades of gray indicate higher probabilities.
      Same parameters are taken as in Fig.~\ref{crossover}.}
    \label{gray}
\end{figure}

Figure~\ref{crossover} shows time evolution of the average number
$n_b$ of molecules (solid line), the superfluid density $|\Phi|^2$
of bosons (dashed line), and the order parameter $|\Delta|^2$
of Fermi pairs (dot-dashed line) per site for (a) $zJ_b/g=0.1$, (b)
$zJ_b/g=1$, and (c) $zJ_b/g=10$ and for $\delta=zJ_b$ and a fixed
$U_b/zJ_b=0.1$. The corresponding probabilities $|f_{n,s}|^2$
for these three regimes are shown in Fig.~\ref{gray}. The initial
molecular ground state is the same in all cases and corresponds to
an almost coherent state.

In the case of weak tunneling, Fig.~\ref{crossover}~(a), the
system dynamics resembles the Jaynes-Cummings limit, see
Fig.~\ref{JCdy} (b): pairs of amplitudes, $f_{n-1,1/2}$ and
$f_{n,-1/2}$, are coupled to each other but uncoupled to the other
amplitudes. This is shown in Fig.~\ref{gray}~(a), where the
probabilities of neighboring states, $(n,sign(s))=(n-1,+)$
and $(n,-)$, oscillate at a frequency $\sim 2\sqrt{n}g$. In
practice, however, each of those amplitudes is of course also
weakly coupled to others due to finite tunneling. Note that the
probability of the $(0,-)$ state remains almost constant at all
times since it has no direct partner.

As $zJ_b/g$ increases, the system dynamics becomes increasingly
dominated by the molecular condensate component. In the extreme
case of $zJ_b/g = 10$, we reach a regime of clear periodic motion
between the molecular and atomic superfluids.
Figures~\ref{gray}~(b) and (c) show that the Jaynes-Cummings character
of Fig.~\ref{gray}~(a) disappears with increasing $zJ_b/g$, and a
high degree of coherence develops, accompanied by the formation of
new order characterized by the simultaneous appearance of several
molecular number states of same spin.

\subsection{Self-trapping transitions}

In this subsection we examine how the dissociation dynamics is affected by the
initial state of the molecular field, or stated another way,
whether it undergoes a sudden transition as a function of
$U_b/zJ_b$.
\begin{figure*}
    \begin{center}
      \includegraphics[height=7.5cm,width=15cm,clip]{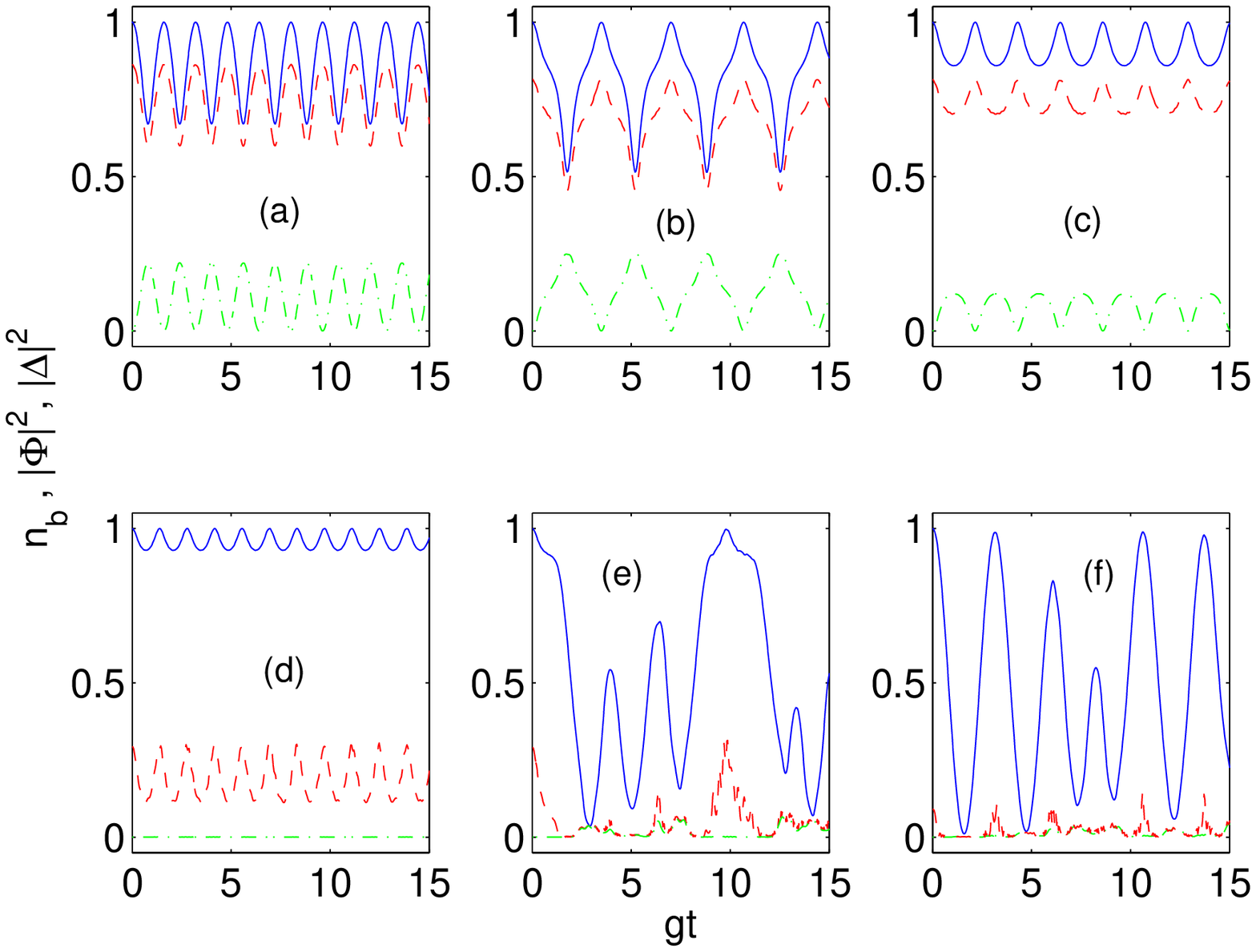}
      \end{center}
    \caption{(Color online) Time evolution of $n_b$ (solid line),
      $|\Phi|^2$ (dashed line),
      and $|\Delta|^2$ (dot-dashed line) for $zJ_b/g=20$, $\delta=0$,
      and for (a) $U_b/zJ_b=1.50$
      (b) $U_b/zJ_b=1.86$, (c) $U_b/zJ_b=1.87$, (d) $U_b/zJ_b=4.70$,
      (e) $U_b/zJ_b=4.71$, and (f) $U_b/zJ_b=5.50$.}
     \label{selftrapping}
\end{figure*}

To answer this question we solved numerically the GW mean-field
equations for several values of $U_b/zJ_b$ for $zJ_b/g=20$ and
$\delta=0$. Fig.~\ref{selftrapping} shows the time evolution of
$n_b$ (solid line), $|\Phi|^2$ (dashed line), and $|\Delta|^2$
(dot-dashed line) from (a) $U_b/zJ_b=1.50$ to (f) $U_b/zJ_b=5.50$.
Since the initial superfluid fraction decreases as $U_b/zJ_b$
increases, these results illustrate situations where the
dissociation dynamics changes from being ``coherent'', or
condensate-dominated, to ``incoherent,'' or non-condensate
dominated. Interestingly, we find two types of transitions
against small variations of $U_b/zJ_b$: the first one is a sudden
suppression of the amplitude of oscillations between
Figs.~\ref{selftrapping} (b), $U_b/zJ_b=1.86$, and (c),
$U_b/zJ_b=1.87$; and the second is a sudden increase of that
amplitude between Figs.~\ref{selftrapping} (d), $U_b/zJ_b=4.70$,
and (e), $U_b/zJ_b=4.71$. Such a transition from a small amplitude
oscillation (localized solution) to a large amplitude oscillation
(delocalized solution), and vice versa, due to a nonlinearity of
the system is known as a self-trapping
transition~\cite{Milburn97,Smerzi97}. Since these two transitions
have very different coherence properties, we refer to them from
now on as the {\it coherent} self-trapping transition for the
former and the {\it incoherent} self-trapping transition for the latter,
respectively.

The Tavis-Cummings model of Sec.~III provides a useful way to
understand why two different self-trapping transitions are present
against a variation of $U_b/zJ_b$. If we regard the third term in
Eq.~(\ref{TCH}), corresponding to the molecule-molecule
interaction, as a nonlinear detuning, the energy difference
between number states of $n$ condensate bosons and $n-1$
condensate bosons is given by $\hbar {\bm \Omega}_n=\hbar\delta_0+\hbar
U_0 (n/N_s)$ where we have neglected a term of order $1/N_s$. The
dynamical properties are determined by the competition between
that detuning  and the photoassociation term in Eq.~(\ref{TCH}).
When the detuning is dominant the dynamics is characterized by
small amplitude oscillations, while large amplitude oscillations
appear when the atom-molecule coupling is dominant. In the
intermediate region where these two effects are comparable, that
is, for ${\bm \Omega}_n \sim g$, a self-trapping transition appears.
Replacing $n/N_s$ in the nonlinear detuning energy ${\bm \Omega}_n$ by the
initial superfluid density of bosons per site, $|\Phi(t=0;U_b/zJ_b)|^2$
which depends on the ratio of $U_b/zJ_b$ as shown in
Fig.~\ref{gssforder}, we obtain, for $\delta=0$ and $U_{bf}=0$,
\[
  {\bm \Omega}_n=-zJ_b+U_b |\Phi(t=0;U_b/zJ_b)|^2
\]
This frequency becomes equal to zero for two values of $U_b/zJ_b$.
For large values of $zJ_b/g$, such as in Figs.~\ref{selftrapping},
self-trapping transitions occur around at these two resonance
points.

Self-trapping transitions can also be understood in a different
picture from the semiclassical form~(\ref{Newtoneq}) of the
generalized Tavis-Cummings Hamiltonian, which describes the motion
of a classical particle in the potential
$V(n_0)$~(\ref{scpotential}). At $t=0$, the particle is at rest
and is located at some point $n_0=|\Phi(t=0)|^2$ on the potential.
Under the influence of the dissociation laser it then starts
rolling down the potential $V(n_0)$ toward smaller values of
$n_0$, passes through a potential minimum, and moves uphill up to
a point having the same potential energy as initially. At this
point the particle turns back, resulting in periodic oscillations
in the (continuous) molecule number. What happens near the
self-trapping condition is that an additional potential barrier
appears in the potential. For delocalized (resp. localized)
solutions, the particle can (resp. cannot) climb up the barrier,
resulting in a sudden transition between large and small amplitude
oscillations. The key factor in achieving this transition is the
quartic form of the potential $V(n_0)$, which disappears in the
absence of the effective two-body interaction
$U_{0}$~\cite{Miyakawa05}.)

\begin{figure}
    \begin{center}
      \includegraphics[height=5cm,width=7.5cm,clip]{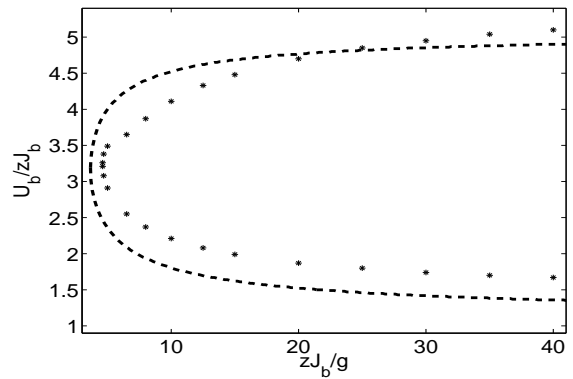}
      \end{center}
    \caption{Critical curve of the self-trapping transitions.
    The star and dashed lines are for the GW mean-field and
    the semiclassical Tavis-Cummings solutions, respectively.
    The inside of the curve is the domain of localized solutions.}
    \label{st_crit}
\end{figure}

This classical interpretation relies on the assumption that the
dynamics of the condensate fraction is uncoupled to that of the
non-condensate component. Since the GW variational state also
involves a non-condensate component, the corresponding mean-field
solutions are expected to deviate somewhat from those of
Eq.~(\ref{Newtoneq}), especially in the vicinity of the {\it
incoherent} self-trapping critical point. Figure~\ref{st_crit}
shows the critical curve of the self-trapping transitions in the
$U_b/zJ_b$ - $zJ_b/g$ plane for $\delta=0$. The dashed line
corresponds to the semiclassical solution of the Tavis-Cummings
dynamics, obtained by replacing $\nu\to \nu_0=|\Phi(t=0)|^2$ in the
potential $V(n_0)$. The stars are mean-field results obtained by
seeking a point where a sudden suppression of amplitude happens in
the first oscillation period~\footnote{For a smaller value of
$zJ_b/g \lesssim 8$, the fluctuations are enhanced, leading to a
situation where large amplitude oscillation shows up after several
small amplitude oscillations. Here, we regard these solutions as
localized (self-trapping) solutions}. Both these results have a
$\subset$ shape, with localized solutions present inside the
curve, and share several common features: first, we observe that
two transitions occur for a given $zJ_b/g$. They correspond to
{\it coherent}, respectively {\it incoherent} transitions, for the
lower and upper values of $U_b/zJ_b$. Second, the transitions
disappear below some value of $zJ_b/g$.

\begin{figure}
    \begin{center}
      \includegraphics[height=5cm,width=7.5cm]{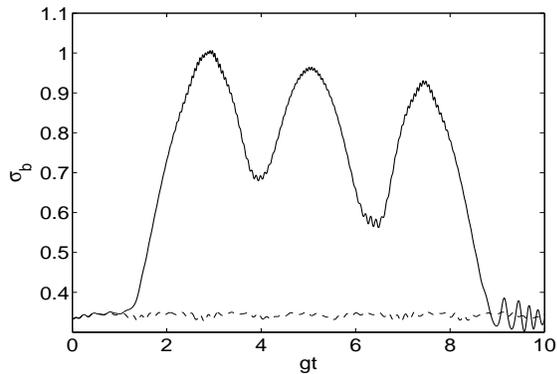}
      \end{center}
    \caption{Molecule number fluctuations per site, $\sigma_{b}$, across the {\it incoherent}
      self-trapping transition point for the cases of Fig.~\ref{selftrapping} (d) (dashed line)
      and (e) (solid line). }
    \label{st_fluc}
\end{figure}

In the vicinity of the {\it incoherent} transition, the mean-field
dynamics of the delocalized solutions is very different from its
semiclassical counterpart. As shown in Fig.~\ref{selftrapping}(e),
a large incoherent oscillation occurs once the molecular
condensate reaches a sufficiently small value, indicating that the
delocalization in the motion of molecular condensate is a trigger
for an incoherent large amplitude oscillation in the system.
This region is also characterized by large fluctuations
\[
\sigma_{b}=\sqrt{\frac{\langle (\hat{n}^b_{i})^2 \rangle
-\langle \hat{n}^b_{i} \rangle^2}{\langle \hat{n}^b_i \rangle}},
\]
in the molecule number fluctuations, a feature that is of course
ignored in the semiclassical description. Fig.~\ref{st_fluc} shows
the time evolution of $\sigma_{b}$ for the cases of
Fig.~\ref{selftrapping} (d) (dashed line), and (e) (solid line).

\subsection{Intersite spin-spin coupling}

Finally, we briefly discuss the effects of XXZ term in the
Hamiltonian $\hat{H}_s$. It is known that the trap frequencies of
homonuclear molecules in a lattice site almost coincide with the
atomic trap frequencies~\cite{Rom04,Ryu05}, so that in general
$J_f \gg J_b$ due to the higher mass of the molecules, $m_b=2m_f$.
However, if the inequality $|U_f|\gg zJ_f$ that underlies all of
the considerations in this paper is fulfilled, the effective
tunneling of the fermions is given by $J_s=4J_f^2/U_f$, which can
then easily lead to a system of $|J_s| \leq J_b$ where the intersite
spin-spin coupling does not affect the coherence properties of the dynamics.
In the following we limit our considerations to this case~\footnote{In the case
of a large $J_s$, an additional self-trapping transition
originating from the intersite spin-spin coupling can be found
even for $U_b=U_{bf}=0$.}. 

In the regime of cooperative dynamics
of Fig.~\ref{mfdpd}, the spin-spin coupling gives rise to a
positive (resp. negative) detuning effect for the case of
attractive (resp. repulsive) fermion-fermion interaction, since a
superfluid state of Fermi pairs ($|\Delta|\neq 0$) lowers, resp.
raises the energy for $U_f<0$, resp. $U_f > 0$. This additional
detuning can however be compensated by changing the
photodissociation detuning energy $\delta$. We have checked that
finite values of $J_s$ preserve the appearance of the
self-trapping transitions. Finally, in the regime of incoherent
dynamics the effects of the XXZ term become weaker with decreasing
$zJ_b/g$. We thus conclude that the discussion in the preceding
subsections holds even in the presence of the intersite spin-spin
coupling.

\section{Summary}

In this paper we have investigated theoretically the
photodissociation of ground-state bosonic molecules into
two-component fermionic atoms in an optical lattice. We showed
that the dissociation dynamics in the strong fermion-fermion
interaction limit, $|U_f|\gg zJ_f$, is governed by a spin-boson
lattice Hamiltonian dynamics. Using a Gutzwiller mean-field
analysis, we examined the crossover of the dissociation dynamics
from an independent single-site regime to a cooperative regime as
the molecular tunneling is increased. We also showed the
appearance of two types of transitions, that is {\it coherent} and
{\it incoherent} self-trapping transitions, as a small variation
of $U_b/zJ_b$. The mean-field approximation only accounts for
those tunnel processes proportional to the expectation values of
bosonic and fermionic pair operators. Thus, when starting from the
MI molecular phase the dynamics of the atoms and molecules inside
the individual lattice sites always remains independent of the
other sites. In this situation, genuine quantum effects are
significant, and are closely related to the stability of the MI
phase under boson-fermion conversion~\cite{Zhou05} and to the
formation of SF correlations across the MI-SF critical
point~\cite{DBHM}. This will be discussed in a future publication.

\vspace{1 cm}

\acknowledgments
This work is supported in part by the US Office
of Naval Research, by the National Science Foundation, by the US
Army Research Office, and by the National Aeronautics and Space
Administration.

%%% REFERENCES %%%%%


\begin{references}
\bibitem{FR} S.~Inouye, M.~R.~Andrews, J.~Stenger, H.-J.~Miesner, D.~M.~ Stamper-Kurn, and W.~Ketterle, Nature (London) {\bf 392}, 151 (1998).
\bibitem{PA} R.~Wynar, R.~S.~Freeland, D.~J.~Han, C.~Ryu, and D.~J.~Heinzen, Science {\bf 287}, 1016 (2000).
\bibitem{Jaksch05} D.~Jaksch and P.~Zoller, Ann.~Phys. {\bf 315}, 52 (2005).
\bibitem{Bloch05} I. Bloch, J.~Phys.~B: At.~Mol.~Opt. {\bf 38}, S629 (2005).
\bibitem{DQPT} J.~Dziarmaga, A.~Smerzi, W.~H.~Zurek, and A.~R.~Bishop,
Phys.~Rev.~Lett. {\bf 88}, 167001 (2002);
B.~Damski, {\it ibid.} {\bf 95}, 035701 (2005);
W.~H.~Zurek, U.~Dorner, and P.~Zoller, {\it ibid.}, 105701 (2005);
J.~Dziarmaga, {\it ibid.}, 245701 (2005); A.~Polkovnikov, Phys.~Rev.~B {\bf 72}, 161201(R) (2005).
\bibitem{DBHM} K.~Sengupta, S.~Powell, and S.~Sachdev, Phys.~Rev.~A {\bf 69}, 053616 (2004);
F.~M.~Cucchietti, B.~Damski, J.~Dziarmaga, and W.~H.~Zurek, e-print cond-mat/0601650.
%\bibitem{Clark04} S.~Clark and D.~Jaksch, {\it ibid.} {\bf 70}, 043612 (2004).
\bibitem{Rom04} T.~Rom, T.~Best, O.~Mandel, A.~Widera, M.~Greiner, T.~W.~H${\rm\ddot{a}}$nsch,
and I.~Bloch, Phys.~Rev.~Lett. {\bf 93}, 073002 (2004).
\bibitem{Kohl} M.~K${\rm\ddot{o}}$hl, H.~Moritz, T.~St${\rm \ddot{o}}$ferle,
K.~G${\rm \ddot{u}}$nter, and T.~Esslinger, Phys.~Rev.~Lett. {\bf 94} 080403 (2005).
\bibitem{Ho06} R.~B.~Diener and T.~L.~Ho, Phys.~Rev.~Lett. {\bf 96}, 010402 (2006).
\bibitem{Stoferle} T.~St${\rm \ddot{o}}$ferle, H.~Moritz, K.~G${\rm \ddot{u}}$nter, M.~K${\rm \ddot{o}}$hl, and T.~Esslinger, Rhys.~Rev.~Lett. {\bf 96}, 030401 (2006).
\bibitem{Busch98} T.~Busch, B.-G.~Englert, K.~Rzazewski, and M.~Wilkends, Found.~Phys. {\bf 28}, 549 (1998).
\bibitem{Blume02} D.~Blume and C.~H.~Greene, Phys.~Rev.~A {\bf 65}, 043613 (2002).
\bibitem{Grimm06} G.~Thalhammer, K.~Winkler, F.~Lang, S.~Schmid, R.~Grimm, and J.~H. Denschlag, Phys.~Rev.~Lett. {\bf 96}, 050402 (2006).
\bibitem{Ryu05} C.~Ryu, X.~Du, E.~Yesilada, A.~M.~Dudarev, S.~Wan, Q.~Niu,
and D.~J.~Heinzen, e-print cond-mat/0508201.
\bibitem{Milburn97} G.~J.~Milburn, J.~Corney, E.~M.~Wright, and D.~F.~Walls, Phys.~Rev.~A {\bf 55}, 4318 (1997).
\bibitem{Smerzi97} A.~Smerzi, S.~Fantoni, S.~Giovanazzi, and S.~R.~Shenoy, Phys.~Rev.~Lett. {\bf 79}, 4950 (1997).
%
\bibitem{Dickerscheid} D.~B.~M. Dickerscheid, U.~Al Khawaja, D.~van Oosten, and H.~T.~C. Stoof, Phys.~Rev.~A {\bf 71}, 043604 (2005).
\bibitem{Carr05} L.~D.~Carr and M.~J.~Holland, Phys.~Rev.~A {\bf 72}, 031604(R) (2005).
\bibitem{Zhou05} F.~Zhou, Phys.~Rev.~B {\bf 72}, 220501(R) (2005);
F.~Zhou and C.~Wu, e-print cond-mat/0511589.
\bibitem{Miyakawa06} T. Miyakawa and P. Meystre, Phys.~Rev.~A {\bf 73}, 021601(R) (2006)
\bibitem{Fisher} M.~P.~A. Fisher, P.~B.~Weichman, G.~Grinstein, and D.~S.~Fisher, Phys.~Rev.~B {\bf 40}, 546 (1989).
\bibitem{Rokhsar91} D.~S.~Rokhsar and B.~G.~Kotliar, Phys.~Rev.~B {\bf 44}, 10328 (1991).
\bibitem{Krauth92} W. Krauth, M. Caffarel, and J-P. Bouchaud, Phys. Rev. B {\bf 45}, 3137 (1992).
\bibitem{Sachdev} S. Sachdev, {\it Quantum Phase Transitions} (Cambridge University Press, New York, 1999).
\bibitem{Jaksch98} D.~Jaksch, C.~Bruder, J.~I.~Cirac, C.~W.~Gardiner, and P.~Zoller, Phys.~Rev.~Lett. {\bf 81}, 3108 (1998).
\bibitem{Greiner02} M.~Greiner, O.~Mandel, T.~Esslinger, T.~W.~H{$\rm\ddot{a}$}nsch, and I.~Bloch, Nature (London) {\bf 415}, 39 (2002).
\bibitem{Emery76} V.~J.~Emery, Phys.~Rev.~B {\bf 14}, 2989 (1976).
\bibitem{Search03} C.~P.~Search, W.~Zhang, and P.~Meystre, Phys.~Rev.~Lett. {\bf 91}, 190401 (2003).
%
\bibitem{Jaksch02} D.~Jaksch, V.~Venturi, J.~I.~Cirac, C.~J.~Williams, and P.~Zoller, Phys.~Rev.~Lett. {\bf 89}, 040402 (2002)
\bibitem{Damski03a} B.~Damski, L.~Santos, E.~Tiemann, M.~Lewenstein, S.~Kotochiqova, P.~Juliene, and P.~Zoller, Phys.~Rev.~Lett. {\bf 90}, 110401 (2003).
\bibitem{Damski03b} B.~Damski, J.~Zakrzewski, L.~Santos, P.~Zoller, and M.~Lewenstein, Phys.~Rev.~Lett. {\bf 91}, 080403 (2003).
\bibitem{Zakrzewski} J.~Zakrzewski, Phys.~Rev.~A {\bf 71}, 043601 (2005).
\bibitem{Fehrmann} H.~Fehrmann, M.~A.~Baranov, B.~Damski, M.~Lewenstein, and
L.~Santos, e-print cond-mat/0307635.
%
\bibitem{Scully} M.~O.~Scully and M.~S.~ Zubairy, {\it Quantum Optics}
(Cambridge University Press, Cambridge, U.K., 1997).
\bibitem{Molmer03} K.~M{\o}lmer, Phys.~Rev.~Lett. {\bf 90} 110403 (2003).
\bibitem{Tavis} M.~Tavis and F.~W.~Cummings, Phys.~Rev. {\bf 170}, 379 (1968).
\bibitem{BCSBECdy} J.~Javanainen, M.~Ko{$\rm\breve{s}$}trun, Y.~Zheng, A.~Carmichael, U.~Shrestha, P.~J.~Meinel,
M.~Mackie, O.~Dannenberg, and K-A.~Suominen, Phys.~Rev.~Lett. {\bf 92}, 200402 (2004); R.~A.~Barankov and L.~S.~Levitov, {\it ibid.} {\bf 93}, 130403 (2004); A.~V.~Andreev, V.~Gurarie, and L.~Radzihovsky, {\it ibid.} {\bf 93}, 130402 (2004);
D.~Meiser and P.~Meystre, {\it ibid.} {\bf 94}, 093001 (2005);
E.~Pazy, I.~Tikhonenkov, Y.~B.~Band, M.~Fleischhauer, and A.~Vardi, {\it ibid.}
{\bf 95}, 170403 (2005).
\bibitem{Uys05} H.~Uys, T.~Miyakawa, D.~Meiser, and P.~Meystre, Phys.~Rev.~A {\bf 72}, 053616 (2005).
\bibitem{Jack05} M.~W.~Jack and H.~Pu, Phys.~Rev.~A {\bf 72}, 063625 (2005).
\bibitem{Miyakawa05} T. Miyakawa and P. Meystre, Phys.~Rev.~A {\bf 71}, 033624 (2005).
\bibitem{elliptic} H.~Hancock, {\it Elliptic Integrals} (Dover, New York, 1917); {\it Handbook of Mathematical Functions}, edited by M.~Abramowitz and I.~A.~Stegun (Dover, New York, 1972).
\end{references}
\end{document}